\documentclass[a4paper,11pt]{article}
\usepackage[english]{babel}
\usepackage{amssymb,amsmath}
\usepackage[hang, nooneline]{subfigure}
\usepackage{epsfig,multicol,ifthen,amsfonts,setspace,graphicx,authblk,fancyhdr,everypage}
\usepackage{graphicx}
 
\begin{document}

\title{\bf{Differential Entropy Dynamics: A Possible Cause of Coherence Resonance }} 

\author[1]{Juhi Rajhans\thanks{rajhansjuhi@gmail.com}}
\author[2]{A.N.Sekar Iyengar \thanks{ansekar.iyengar@saha.ac.in}}

\affil[1]{Department of Physical Sciences, Indian Institute of Science Education and Research- Kolkata, India}

\affil[2]{Plasma Physics Division, Saha Institute of Nuclear Physics, Kolkata, India}

\pagestyle{fancy}

\maketitle

\thispagestyle{plain}
\section{\bf{Abstract}}

Coherence resonance can be explained using differential entropy and mutual information. This theory explores the role of external noise in stabilising chaotic circuits such as the uni-junction transistor relaxation oscillator.The phenomenon of coherence resonance maximizes differential entropy and mutual information.Thus most natural chaotic oscillators show coherence resonance in the presence of an external driving noise.

\section{\bf{ Keywords}}
Coherence resonance, differential entropy, mutual information .

\section{\bf{Introduction}}

Electronic circuits form the most elegant and simplest models to study synchronisation in chaotic oscillator systems.One of the most primitive models is the uni-junction transistor in the threshold voltage limit.A uni-junction transistor (UJT) is a bar of n-type semiconductor with a pinch of p-type semiconductor in the middle. It is a two base one emitter transistor with current flowing in the base2-emitter-base1 direction. Thus the CRO shows the voltage peaks in the emitter-base1 and emitter-base2 junctions of opposite polarity. Time dependent external stochastic perturbation leads to large amplitude limit-cycle oscillations in the emitter of the uni-junction transistor.According to Nurrujaman[1], thermal noise and negative resistance across the emitter and one of the bases makes it a non-linear relaxation oscillator. The phenomenon of conductivity modulation in the emitter-base1 region is responsible for the negative resistance of the emitter-base1 junction.\\ \\

 The central theme of this paper is to explain coherence resonance in period-one limit cycle oscillator like the UJT.The study shows that synchronisation of noisy semiconductor circuits in the presence of external white Gaussian noise is inevitable.
  
\section{\bf{Theory and Calculations}}

\subsection{\bf{UJT characteristics without external noise}}
The crucial point is the onset of negative resistance characteristics.Negative resistance occurs at zero frequency when $\frac{dV}{dI} < 0 $. The conductivity of a semiconductor can be altered by the injection or extraction of carriers. Nishi[2] shows in his work that negative resistance under D.C voltage can be obtained under two conditions-\\
(1)In one model the electrons and holes in the emitter-base1 junction  show Brownian motion with drift and no recombination( of electrons and holes) and in the other\\
(2) where the lifetime of the drifting electrons and holes increases with increasing carrier density.\\



Calculating the values of peak and valley voltages in terms of the power supply voltages, and then calculating the time required to discharge to these
voltages, we obtain a set of equations.
\begin{center}

$  \frac{dV}{dt} = \frac{V_{0}}{(RC)}(1- \exp{\frac{-t}{RC}})$ if $ t \leq RC(1+3i)$

$   \frac{dV}{dt} = - \frac{V_{0}}{RC}\exp{(-\frac{t^{x}}{RC})}$ if $ 2iRC < t \leq 3iRC$

    $ i=0,1,2,3,4,.... ,  1<x <2 $.
    
\end{center}    

The biasing voltage determines the dynamics at the emitter- base1 junction. The dynamics in this region is null till the threshold voltage is crossed following which limit-cycle relaxation oscillations set in. This suggests the formation of a saddle-node bifurcation which leads to the formation of a stable and an unstable fixed point.In order to study the dynamics, one has to consider a 1-D graph with similar characteristics. The simplest such graph is the      map governed by the following relation.\\
\begin{center}

$ x_{n+1} = ax_{n}$ for $0\leq x_{n} \leq \frac{1}{2}$\\
$ x_{n+1} = a(1 - x_{n})$ for $\frac{1}{2}\leq x_{n} \leq 1 $\\
\end{center}

Thus, according to the graph the limit cycle occurs for a certain value of the parameter and then the attractor goes through rapid period doubling into chaos. The brief window of parameter values which sustains the limit cycle is when the stable fixed point exists. The limit cycle collapses when the unstable fixed point meets the stable manifold leading to internal crisis.

\subsection{\bf{UJT characteristics with \\ external noise}}

\subsubsection{ Stable oscillations due to AWGN resonance}

It has been proposed that conductivity modulation generates a Johnson-Nyquist noise in the semiconductor base.This implies that the thermal noise generated due to mobile electrons is additive white Gaussian (AWGN) in nature,(Wong[3]).\\

The internal Johnson-Nyquist noise and external stochastic  AWGN fed into the emitter through a capacitor, interfere and generate another white Gaussian noise. $X_i = N(0,n1)$ is a Gaussian with zero mean and variance n1. $Y_i = N(0,n)$ is another Gaussian noise . Thus, from the theorem of addition of two probability distribution functions, $ Z_i = X_i +Y_i $ is also a white Gaussian noise.The addition of external Gaussian noise acts as a perturbation and throws the system into the stable limit cycle, and the amplitude increases due to differential entropy and mutual information maximisation. Following is a proof of the above statement(Reza[4]). \\

The differential entropy of a Gaussian is given by \\
\begin{center}

$ h(X) = -\int_{-\infty}^{+\infty}f_{X}(x)\log(f_{X}(x))$\\
$ = -E[\log(f_{X}(x)] $
\end{center}
Now, we choose to maximize the differential entropy. \\
\begin{center}
$ \int p(x)dx = 1 $\\ $ \int x p(x)dx = \mu $\\ $ \int (x-\mu)^{2} p(x)dx = \sigma^{2} $\\
\end{center}
Using the Lagrangian multiplier one finds the following functional-\\
\begin{center}
$  F = -\int p(x)\ln p(x)dx + \lambda_{1}(\int p(x)dx - 1) + \lambda_{2}(\int x p(x)dx - \mu) + \lambda_{3}\int (x - \mu)^{2}p(x)dx - \sigma^{2})$\
\end{center}
From the calculus of variations,
\begin{center}
$  p(x) = exp{ -1 + \lambda_{1} + \lambda_{2}x + \lambda_{3}(x - \mu)^{2}}$\\
$ p(x)= \frac{1}{2 \pi\sigma^{2}}^{\frac{1}{2}}\exp(-\frac{(x - \mu)^2}{\sigma}^{2} $
\end{center}
The value of the maximum entropy-
\begin{center}
 H(x)= $ \frac{1}{2}(1 + ln(2\pi\sigma^{2})) $
 \end{center} 
 Similarly, the mutual information can be written down in the following form-
\begin{center}
 $ I[x,y] = \int\int p(x,y)ln(\frac{p(x)p(y)}{p(x,y)})dxdy $
 \end{center}
 One could also show that the mutual information is maximised when p(x), p(y) is a Gaussian.
 \begin{center}
 $ I[x,y] = H[x] - H[x|y] = H[y] - H[y|x] $
 \end{center}
 Thus maximizing the mutual information is equivalent to maximizing H[y]. The previous calculations show that this is possible only when y is circularly symmetric complex Gaussian or white Gaussian(Telatar[5]) and this happens when x is circularly symmetric complex Gaussian. The maximal mutual information is given by the following- \begin{center}
 $ I[x:y] = \log{\det{(I + HQH^{\dagger})}} = \log{\det{(I + QHH^{\dagger})}} $
\end{center}  
where the equality follows from the determinant identity $ \log{\det{(I + AB)}} = \log{\det{(I + AB)}}  $. H is a matrix relating the signals x and y. Q is the covariance of x and y. \\ \\
 
 Mutual information is the reduction in uncertainty about x given a value of y or vice-versa. T One can interpret the reduction in uncertainty in the second signal as onset of periodicity of the signal after superposition with the external noise.[8]\\

However, the question still remains as to why limit cycle oscillations should begin at a particular amplitude of the external noise and not at lower amplitudes.
Silvia de Monte[6] shows in her work that low amplitude of noise to a system at a stable fixed point or very close to it leads to bifurcation.The system traverses the entire
phase space before returning back to the fixed point when the noise amplitude is equal to the emitter voltage at no noise-relaxation-oscillations. 

\section{\textbf{Coupled UJTs in the presence of external noise- Coherence Resonance }}

The differential entropy analysis can be extended similarly with two or more UJTs coupled to each other in the presence of an external white Gaussian noise. The differential entropy of each such oscillator is maximum in the resonance regime. \begin{center}

$ I[x;y;z] = H[x] + H[y] + H[z] - H[x|y] - H[y|z] - H[z|x] $

\end{center}

where one of the signals is the external noise and the other two are the oscillator signals. It is obvious that I[x;y;z] is maximum when H[x], H[y], H[z] is maximum, i.e, each of them generates a circularly symmetric complex signal. This would generate a shifted Arnold tongue diagram which can be tested through experiments.

\section{\textbf{Conclusions}}

Thus, the algebra of differential entropy and mutual information can be extended to coupled oscillator systems in the presence of an external noise. Numerous studies show that noise-induced resonance is observed in various biological oscillators. The explanation through differential entropy can be directly extended for more number of oscillators and the Arnold tongue diagram would suggest different mode-locked orbits in presence of the noise. We conclude by saying that the evolution of differential entropy and mutual information in chaotic oscillators are clearly suggestive of emerging patterns in complex systems. We also extend our heartfelt thanks to Dr. S.K. Dana whose valuable insights and discussions at Indian Institute of Chemical Biology gave us a crucial paradigm to understand synchronisation in a chaotic system like the UJT.

\end{document}